# AI-in-the-Loop Planning for Transportation Electrification: Case Studies from Austin, Texas

Seung Jun Choi
Urban Information Lab
The University of Texas at Austin
Austin, Texas, USA
jun.choi@utexas.edu

*Abstract*—This study explores the integration of AI in transportation electrification planning in Austin, TX, focusing on the use of Geospatial AI (GeoAI), Generative AI (GenAI), and Large Language Models (LLMs). GeoAI enhances site selection, localized GenAI models support meta-level estimations, and LLMs enable scenario simulations. These AI applications require human oversight. GeoAI outputs must be evaluated with land use data, GenAI models are not always accurate, and LLMs are prone to hallucinations. To ensure accountable planning, human planners must work alongside AI agents. Establishing a community feedback loop is essential to audit automated decisions. Planners should place Community Experience (CX) at the center of Urban Planning AI.

*Keywords—urban planning AI, transportation electrification, electric vehicle, electric vehicle charging, community experience*

## I. Introduction

The rise of Electric Vehicles (EVs) has introduced the transportation electrification plan. This plan focuses on the installation and management of infrastructure needed to support EVs, including charging stations, grid coordination, and electricity assessments, often involving multiple stakeholders through public-private partnerships. A common approach to identifying suitable locations for public EV chargers is site suitability analysis using GIS [1], [2].

With the emergence of concerns about skewed EV adoption and uneven public charger deployment [3], issues of distributive fairness in EV infrastructure planning persist. During the Biden administration, the USDOE's Argonne National Laboratory conducted site suitability analyses with a focus on equity. Their analytical framework aligned with the federal Justice40 initiative, which mandates that 40% of government investments benefit Disadvantaged Communities (DACs). To support this effort, tools such as the Climate and Economic Justice Screening Tool and the EV Charging Justice40 Map Tool were developed. These maps identify census tracts with relatively limited access to public EV charging stations.

Meanwhile, the rise of Artificial Intelligence (AI) has been transforming planning practices. Peng et al. [4] coined the term Urban Planning AI (Urban AI) to describe the use of AI applications or agents, either partially or fully, throughout the planning process. The degree of human planner involvement varies depending on how AI is used. Urban AI expectations and challenges differ by discipline.

Setting aside debates over whether the use of AI in the planning process is right or wrong, this research focuses on how AI can be properly applied. Recognizing that contemporary planning efforts are mainly grounded in normative frameworks and equity-oriented approaches [5], the analytical framework within the realm of Urban AI is rooted in principles of equity/justice. This research presents multiple case studies illustrating how AI applications and AI agents can complement the transportation electrification planning process. These include Geospatial AI (GeoAI), Generative AI (GenAI), and Large Language Models (LLMs). By synthesizing the expected outcomes and associated challenges of each AI type, this study identifies the level of human oversight required when integrating AI into planning processes. This paper explores the following research questions:

1. What are the expected benefits, challenges, and outcomes of applying Urban AI to transportation electrification planning?

2. How should AI be appropriately integrated into the planning process, and what level of human oversight is necessary, and when?

Following the introduction, the next section discusses the use of different types of AI in the planning process from both positive and normative perspectives. The present study then presents the methods and materials used in this study, followed by the analysis results and corresponding discussion. Finally, the study concludes by synthesizing the analytical findings and discussion to offer guidance for future applications.

## II. Literature review

### A. AI and the Planner: Adapting to a New Technological Landscape

Brooks [5] categorized planning as a distinct field of research and practice, characterized by its future-oriented nature, its advocacy for public goods, and its engagement in normative dialogues about what society ought to be. Planning practice can be either positive or normative. Positive planning is comparable to the work of natural scientists, where hypotheses are tested to explain how things operate. In contrast, normative planning involves the application of external principles, such as equity, justice, and fairness, or aligns with political goals or value-based reasoning, such as in rational planning aimed at achieving desirable outcomes.

Since the 1970s and 1980s, planners have been introduced to computational tools, and by the 21st century, computational modeling, including large-scale urban models, rule-based models, cellular automata models, agent-based models, and GIS, has become the dominant paradigm among positivist planners [4]. Today, AI applications such as machine learning, deep learning, and data mining techniques, as well as AI agents like computer programs, software tools, and LLMs, are bringing new dimensions.

Planners will soon be, or are already being, asked to address issues for which they have little to no prior experience. The advent of GenAI and LLMs is beginning to replace some of the routine tasks traditionally carried out by planners and urban designers. Models such as DALL·E, Midjourney, Gemini, DreamStudio, and Adobe Firefly can now create visuals that align with planners' intentions. OpenAI's Sora further advances this capability by generating real-time video scenarios from either text or image inputs. With the reasoning capabilities now embedded in LLMs and the recent release of OpenAI's Deep Research, academia is undergoing a transition in how knowledge is produced.

At a minimum, partial knowledge of AI and information technology is now required across nearly all professions. Contemporary and future planners should be interdisciplinarily equipped, possessing both computational expertise and a strong foundation in planning. The emergence of higher education programs focused on computational skills and urban technology within planning schools, as well as the introduction of courses on AI applications, serves as evidence of this new normal.

*B. Use Cases of AI in Positive Planning*

Peng et al. [4] categorize Urban AI into four types: assisted, augmented, automated, and autonomized planning. Each category represents empirical applications of AI tools and agents in planning practices. Although these classifications often overlap and are not strictly separated, each type reflects varying degrees of AI involvement and different roles for planners within the planning process.

From a broader perspective, the use of Urban AI, whether to explain how systems operate through computational modeling or to generate new datasets and AI-driven insights, is fundamentally empirical and aligns with positivist planning approaches. For example, Jang et al. [6] employed LLMs to analyze place identity based on specific criteria. Quan [7] developed a generative model capable of producing urban forms to augment urban design practices across different cities. Computer vision applications help uncover urban complexities by enhancing our understanding of the natural environment, human settlements, and domain-specific planning [8].

Similar to planning scholars, planning practitioners with computational knowledge can develop their own models. Building localized forecasting or classification models can help predict desired outcomes within a supervised learning framework. This approach can be extended further by incorporating explainable machine learning methods to investigate the underlying causes of those outcomes. Unsupervised learning can be used to profile communities, and the labels generated through clustering can later be reintegrated into a supervised learning workflow. Reinforcement learning offers another avenue, enabling practitioners to simulate scenarios and experiment with different computational settings, such as agents, policies, rewards, and penalties. These simulations require empirical validation to ensure real-world applicability. Additionally, using LLMs to optimize infrastructure plans, summarize documents, and support public engagement is a growing trend in planning practice.

However, not everyone possesses the same set of skills and knowledge. Some users exhibit reluctance toward the use of AI in practice. The gap between theory and practice is inevitable, and in fact, beneficial, as it fosters intellectual self-reflection [9]. For practitioners with limited computational expertise, using Graphical User Interface (GUI)-based tools remains a practical option. For instance, ESRI's ArcGIS Pro provides deep learning toolkits and pretrained models that support tasks such as image feature extraction, pixel classification, object tracking, and image redaction. Many publicly available pretrained models, featured on platforms like GitHub or Hugging Face, can be leveraged to support planning work without the need to build models from scratch.

*C. Use Cases of AI in Normative Planning*

Using AI in normative planning involves not only computational modeling to promote certain societal values but also guiding Urban AI applications in alignment with the core responsibilities of planners, such as advancing public goods, advocacy, equity and justice, communicative action, and anticipating information and misinformation. It also requires ongoing dialogue about how AI should be used in planning.

Some computational works explicitly aim to support disadvantaged groups and contribute to discussions around equity and justice. For example, Choi and Jiao [10] developed an interactive dashboard with forecasting capabilities to identify transit service gaps. Rojas et al. [11] implemented a machine learning framework with continuous human oversight during deployment to promote health equity. Other studies emphasize the integration of participatory planning practices to enhance the accessibility and distribution of computational insights, helping to shape how AI-driven decisions should be shared with the public [12]. These efforts intersect with work from computational and information science disciplines. Principles from human-centered design, ethical AI guidelines, and Human-AI (HAI) ethics, encompassing transparency, fairness, security, accountability, power, trust, privacy, explainability, and inclusion, are essential to the normative integration of AI in planning practice. The interpretation of each element may vary based on whether it is intended for profit or for non-profit objectives [13].

Empirical models are successful when they have a meaningful impact on communities. Calling for using AI normatively in planning process and practice would require qualitative aspects by conducting survey, focus groups, interview, experiment, data audit, observation, or rely on ground theory. Yet, much of the work on Urban AI remains primarily focused on empirical aspects. Boeing et al. [14] raised concerns about the neglect of critical issues such as privacy, ethics, and, ultimately, human dignity within the knowledge domain of urban analytics. Peng et al. [4] argued that both planning academia and professional practice are not adequately prepared to address the potential pitfalls of AI adoption. Planning schools, therefore, should strive to offer a dual approach: one that equips students with the skills to build computational models and use AI tools, and another that fosters the ability to critically situate values within AI applications and the deployment of AI agents.

III. METHODS AND MATERIALS

*A. Study Area*

The study area is Austin, TX, including three counties, Travis, Hays, and Williamson, that lie along the official EV corridor, I-35 (as defined under Section 1413 of the FAST Act). These counties were selected because they exhibit challenges related to distributive fairness in public EV Charging Station (EVCS) allocation and EV adoption [15]. The National Argonne Laboratory suggests classifying regions into three categories, Transportation Network Company (TNC), corridor, and rural areas, and conducting site suitability analyses separately for each [2]. Following their guidelines, we categorized census block groups accordingly. **Figure 1** presents the classification of census block groups (see **Figure 1a**) and the locations of DACs designated by Justice40 (see **Figure 1b**) within the study area. Rural areas were defined as low-density regions with 200 or fewer housing units per square mile [16]. Corridor areas were identified as those located within one mile of an EV corridor. All remaining areas were classified as TNC. Among the 1,206 census block groups in the Austin area, 693 (58%) were classified as TNC, 382 (31%) as corridor, and 131 (11%) as rural.

*B. Materials*

The site suitability analytical framework is based on the work of Argonne National Laboratory [2]. Key factors considered include population density, traffic volume, existing EVCS locations, racial composition, income levels, energy and grid infrastructure, land use, housing units, vehicle ownership, proximity to EV corridors, and policy factors. Relative access to public EVCS can be assessed across population density and various sociodemographic factors [17], [18]. DACs, such as low-income populations and communities of color, often face limited access to public EVCS [3], [19]. Compared to urban settings, rural areas tend to have significantly lower access to public EVCS [15]. Built environment characteristics [20] and the availability of public EVCS [21], [22] also play a critical role in influencing EV adoption.

Sociodemographic factors were based on the 2018–2022 ACS 5-Year Estimates obtained through Social Explorer. Population density was measured as the number of residents per square mile. The race variable reflected the proportion of Hispanic and Black populations, while the income variable captured the proportion of households below the poverty threshold. The housing variable considered the share of multifamily (MF) housing units. Vehicle ownership was measured by the percentage of zero-vehicle households. Traffic density was calculated using the highest average annual daily traffic, with values preprocessed using Euclidean distance in ArcMap. Traffic data for 2024 were obtained from the Texas Department of Transportation.

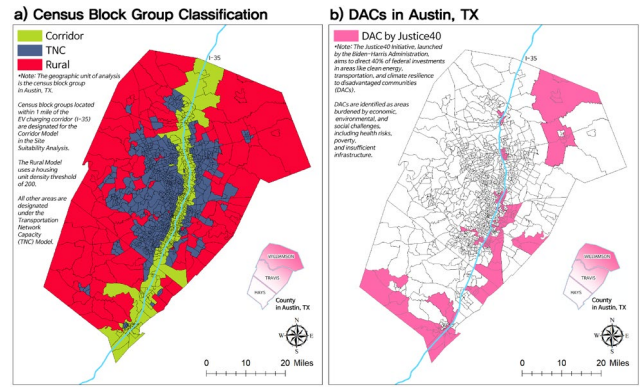

**Fig. 1.** Study Area for Site Suitability Analysis: (a) Census Block Gorup Classification Result; (b) Justice40's Disadvantaged Communities

The list and addresses of EVCS locations were retrieved using NREL's API. Level 2 and Level 3 chargers were classified as public EVCS, and their distances were also processed using Euclidean distance methods. The nearest distance to a non-Tesla DCFC was also considered. The energy/grid variable measured the shortest distance to substations, specifically those with a minimum capacity of 110 kV and 220 kV. The EV corridor variable was derived from the nearest distance to I-35. A dummy variable representing DACs, as designated by the Justice40, was included as a policy factor.

GeoAI refers to spatially explicit AI that analyzes multi-source spatial data and incorporates locational information [23]. The land use factor was derived from GeoAI's land cover classification estimates. Land cover was categorized into four types: developed, barren, green/forest, and water. A Landsat 9 image with less than 5% cloud cover from December 2024 was obtained from the USGS. Land cover classification was performed using a pre-trained supervised Support Vector Machine model in ESRI's ArcGIS Pro, based on a sample size of 100.

All collected variables were preprocessed using quantile normalization for raster data. In general, higher values were assigned greater priority, except in cases where proximity was the key criterion, such as the nearest distance to substations or to I-35, where smaller distances were prioritized. **Figure 2** presents the preprocessed geospatial layers used in the base site suitability analysis. Descriptive statistics for all collected variables are provided in **Appendix Table A**.

Additionally, this study employed a GeoAI-based parking lot classification model to augment the base site suitability analysis. The model is offered by ESRI and fine-tuned for the U.S. context. It was applied to high-resolution imagery with a spatial resolution of 50 centimeters, using a total of 70 images. To mitigate inaccuracies in the classification results, estimated parcels located within private residential areas or environmental amenities were excluded, in alignment with land preservation efforts. Due to the limited availability of zoning regulation data, the classification and filtering were restricted to areas within the Austin City Limit. The land use map was based on the City of Austin's 2023 dataset.

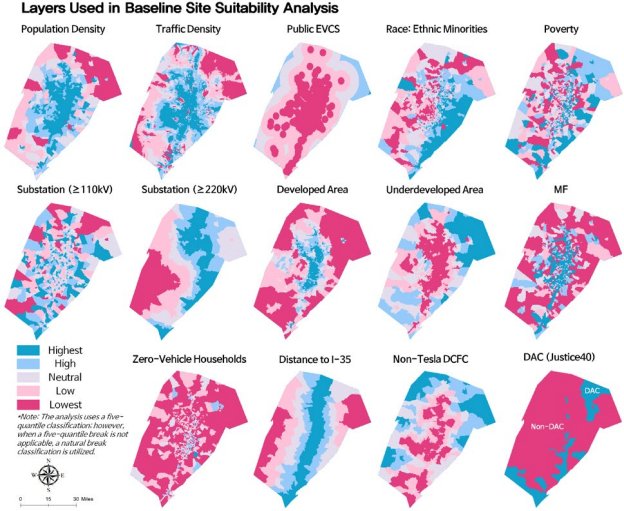

**Fig. 2.** Layers Used in the Baseline Site Suitability Analysis in Austin, TX

For the most up-to-date modeling practices, GenAI modeling was implemented using the Pix2Pix GAN. The model was trained on manually created paired samples consisting of a Landsat image and the corresponding base site suitability analysis result. Each training pair included two images of 256×256 pixels, combined into a single 512×256 pixel input. For site scenario planning, the most suitable locations were identified using Google Earth Pro, from which 3D tiles were extracted. Publicly available 3D meshes representing chargers, stations, and parking spaces were then overlaid onto the site. OpenAI's Sora was used to generate real-time footage and visualize the site plan. The chatbot interface was designed using Voiceflow and integrated with ChatGPT. The computational modeling was supported by an NVIDIA® RTX™ 6000 Ada Generation GPU.

*C. Methods*

The case studies consist of three research phases. First, a base site suitability analysis was conducted. The TNC, rural, and corridor areas required different sets of variables for analysis [2]. Each area underwent a separate site suitability analysis, and the results were synthesized using an image analysis tool in GIS. Suitability analysis outcomes were categorized into four levels, ranging from most suitable (Level 1) to least suitable (Level 4).

The TNC model considered population density, traffic density, availability of public EVCS, proximity to substations with a minimum capacity of 220 kV, developed land use, racial composition, income level, housing type, and vehicle ownership. The rural model included population density, traffic density, public EVCS availability, distance to substations with at least 220 kV capacity, and underdeveloped land use. The corridor model focused on population density, traffic density, proximity to substations with at least 110 kV capacity, designation as a Justice40 DAC, distance to the EV corridor, and distance to non-Tesla DCFC.

Second, computational results, including outputs from GeoAI, the revised site suitability analysis, and GenAI, are presented. The GeoAI component includes land cover classification and parking lot classification results. The updated site suitability analysis was conducted based on the initial base site suitability framework, with separate models for TNC, rural, and corridor areas. Each model was further refined using the parking lot classification outcomes. For the GenAI modeling, a total of 103 samples were used, with 80% allocated for training and 20% for testing. Hyperparameter settings followed those in the original Pix2Pix GAN paper by Isola et al. [24]. The buffer size was set to 200, the batch size to 2, and the number of training epochs to 200. At the final epoch, the average Peak Signal-to-Noise Ratio (PSNR) and Structural Similarity Index Measure (SSIM) were calculated using both training and testing samples. These metrics evaluate the similarity between the predicted output and the ground truth images. Higher values of PSNR and SSIM indicate better reconstruction quality. SSIM values range from 0 to 1.

Third, a site scenario plan is presented. Representative samples selected from the site suitability analysis, augmented by GeoAI results, are visualized in 3D. The potential installation of public EVCS is illustrated using 3D objects. OpenAI's Sora was used to generate scenarios of EVCS installation; however, the model was not informed whether each site represented a TNC, rural, or corridor area. The chatbot interaction includes user greetings, query confirmation, progress tracking, and site review. A welcoming image of a robot was generated using DALL·E. ChatGPT was integrated to generate responses.

## IV. RESULTS AND DISCUSSION

*A. Contemporary Site Suitability Analysis Results*

**Figure 3** presents the results of the base site suitability analysis. The individual analyses for the TNC, rural, and corridor models are shown in **Figure 3a**, while the synthesized result is displayed in **Figure 3b**. The TNC model identified areas near the corridor as the most suitable, whereas West Austin and Northeast Austin were considered the least suitable. The rural model found East Austin to be relatively more suitable than West Austin. The corridor model primarily prioritized areas adjacent to the corridor. When synthesized, the combined analysis confirmed that regions near the corridor are the most suitable overall, with East Austin rated as more suitable than West Austin.

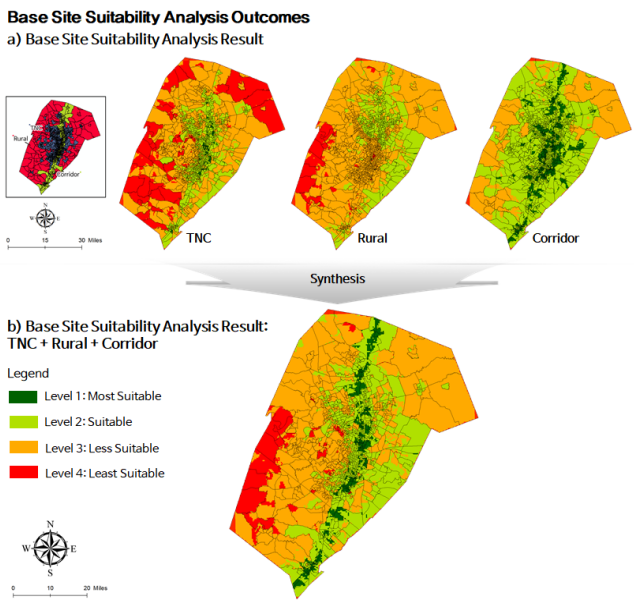

Fig. 3. Base Site Suitability Analysis Results: (a) Results from Separate Analyses; (b) Synthesized Result Integrating TNC, Rural, and Corridor Models

The prioritization of corridor sites has previously been critiqued. Carlton and Sultana [25] pointed out a grey area overlooked by the Justice40 framework, where the National Electric Vehicle Infrastructure (NEVI) program was fully, and the Charging and Fueling Infrastructure (CFI) Discretionary Grant Program was primarily, focused on alternative fuel corridors for charger installation, leading to issues of overconcentration. In the U.S., census tracts located near corridors are likely to have 2.7 times more overall public chargers and 5.3 times more fast chargers [25]. This gap is even more pronounced in rural areas than in urban ones [15], [25]. The limitations of state-centric programs underscore the shortcomings of a reformist orientation in equity analysis, emphasizing the need for community-led and justice-driven alternatives [25], [26]. The Justice40 initiative is no longer continued under the Trump administration, it remains unclear whether corridors will continue to be prioritized for charger installation.

### B. AI-augmented Site Suitability Analysis: Computational Estimation Results

**Figure 4** presents the estimation results generated by GeoAI. In **Figure 4a**, a LandSat image was classified into land cover categories, with Central and Downtown Austin predominantly identified as developed land. The land cover classification results were used for the base site suitability analysis. **Figure 4b** provides sample illustrations of parking lot classification. The model effectively avoided making predictions over green open spaces while accurately preserving building structures, as visible from the bird's-eye view. **Figure 4c** displays consolidated parking lot parcels within the Austin city limits, filtered according to land use zoning regulations. Most vacant parking lots were found in Central Austin. **Figure 5a** presents the site suitability analysis augmented with GeoAI estimations. It compares the most and least suitable areas identified by both the base analysis and the GeoAI-augmented analysis. Overall, the spatial extent of both the most and least suitable regions decreased when incorporating GeoAI estimations. However, the GeoAI-based analysis identified several distinct sites. Although the magnitude of change is relatively small, some additional areas were identified as the most suitable land, while a small region was newly identified as the least suitable.

**Figures 5b** and **5c** present the GenAI modeling results and estimations using the Pix2Pix GAN for samples from the training and testing datasets. The model was more effective at translating LandSat images into suitable land predictions. However, it was less accurate in identifying the least suitable land, as observed in the training samples. In the testing samples, the model provided approximate proxy measures but failed to fully capture suitability relative to the ground truth in some cases. In terms of evaluation metrics, the average PSNR for the training set was 22.8, and the SSIM was 0.87. Performance declined in the testing set, with an average PSNR of 17.5 and an SSIM of 0.82.

Based on the analytical results, GeoAI augments the analytical framework by enabling comprehensive land cover classification and identifying parcels that may otherwise be omitted. Since a complete land use map was not available within the Austin City Limit, supervised classification using GeoAI was employed to determine land use characteristics and general categories. The site suitability analysis, when augmented with GeoAI, helped narrow the scope and identify distinct areas with high suitability. Moving forward, with the application of GenAI modeling, planners can automate processes such as estimation, which traditionally required manual data collection, preprocessing, and analysis. Similar to how meta-analyses, such as Ewing and Cervero [27], provide average coefficients across studies, generative pre-trained models can serve a comparable role. Releasing such models could benefit rural regions or tribal nations with limited research resources by enabling proxy estimations.

It is important to acknowledge that no empirical model is perfect. Human oversight remains essential in the application of both GeoAI and GenAI. In this study, GeoAI estimations were evaluated prior to use based on existing land use information. There were also instances where the Pix2Pix GAN model failed to produce reliable outputs. A truly fair or equitable estimation by GenAI would require further evaluation, including performance testing across different community groups. Auditing estimations and having human planners verify AI outputs in the field are critical steps. While automated systems may eventually enable AI to lead parts of the planning process [4], it is human planners, those who model, design, and implement these tools, who remain at the forefront. The responsibility for managing, updating, and debugging such models ultimately lies with us.

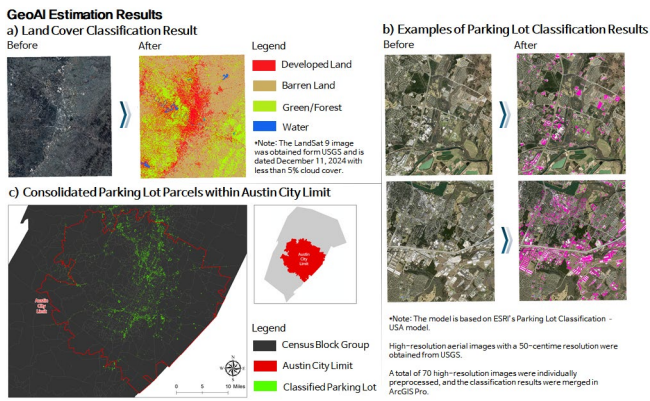

**Fig. 4.** GeoAI Estimation Results: (a) Land Cover Classification; (b) Examples of Parking Lot Classification; (c) Consolidated Parking Lot Parcels within the Austin City Limit

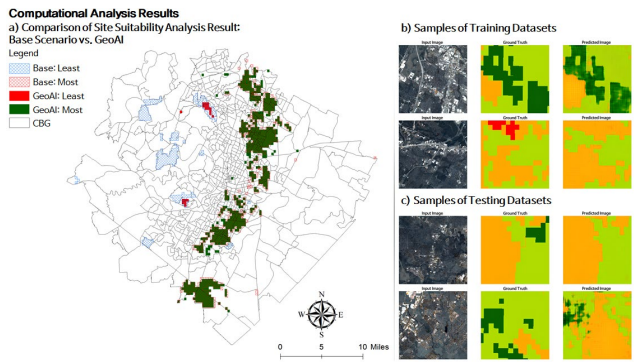

**Fig. 5.** Computational Analysis Results: (a) Comparison of Site Suitability Analysis Result; (b) Samples of Training Datasets in Pix2Pix GAN; (c) Estimation Samples of Testing Datasets in Pix2Pix GAN

*C. Site Scenario Plan*

**Figure 6** presents the computational application of AI in site planning. Representative sites from TNC and corridor areas were selected; no highly suitable site was identified in rural areas. 3D tiles were obtained from Google Earth Pro, and with human oversight, planners can overlay 3D objects for scenario planning (see **Figure 6a**). For the TNC site, additional chargers were placed in parking lots near a mega-retail store. In the corridor site, extra parking spaces or charging stations were added. **Figure 6b** shows images and visual outputs generated by OpenAI Sora using the same base images as in the human-oversighted plans. Sora is capable of editing or adding desired elements to site plans. For the TNC site, Sora envisioned adding parking spaces and EV chargers with solar panels. For the corridor site, it recommended installing additional chargers, hubs, or stations near the interstate highway intersection. Sora can also simulate vehicle movement, human behavior, and projected outcomes based on user-defined requests.

**Figure 6c** illustrates the chatbot interface and its demonstration, designed to assist planners in managing site plans. The interaction begins with a welcoming message to the user. The user submits a query, and the model responds with confirmation. Using ChatGPT, responses can be paraphrased in various forms to enhance communication. The chatbot then progresses to the next step by asking the planner to specify the target site type (e.g., TNC, corridor, or rural). Upon selecting the TNC site, the model proceeds to review the current site plan and charger operations.

Currently, empirical research on digital twins is gaining significant attention. In urban planning, most applications focus on 3D visualizations [28], [29], [30]. Existing studies often justify digital twin applications by proposing imaginative corrections to empirical estimations. However, a true digital twin of a city should simulate how people actually behave in urban environments. While there is ongoing debate about whether 3D visualization alone qualifies as a digital twin, the development of open-source tools and software (e.g., Google Earth Pro, Blosom, Leafmap, Cesium, Nvidia's Omniverse, and Cosmos) has undoubtedly made it easier to create digital replicas of physical settings. GenAI models have been developed to convert 2D images into 3D or generate 3D objects directly, significantly reducing the time required for designers. These advancements enable planners to experiment with scenario-based site plans more efficiently.

Surprisingly, Sora was capable of generating site plans based on a given set of inputs. Although it may require multiple iterations to achieve the desired outcome, its ability to produce not only site layouts but also dynamic videos with simulated human and vehicle movements is remarkable. As AI models and fine-tuned agents for site planning and urban design continue to advance, related professions may face significant transformation. Of course, AI has a known tendency to produce hallucinations, and there is some resistance to this characteristic. Interestingly, Yao et al. [31] challenge the negative framing by arguing that hallucination in LLMs is a feature rather than a bug, comparable to the creativity found in human agency. Whether we choose to accept AI-generated suggestions may ultimately depend on a kind of human Turing test. If hallucination cannot be fully eliminated, we may need to tolerate a certain level of it. The real issue then becomes determining the appropriate confidence threshold for trusting AI-generated decisions.

In the meantime, integrating LLMs with urban infrastructure, systems, and physical forms has become increasingly popular. Virtual assistants powered by LLMs can leverage extensive digital knowledge to support planners in a complementary way. OpenAI outlines five stages of Artificial General Intelligence (AGI): chatbots, reasoners, agents, innovators, and organizers [32]. The localized chatbot demonstration in this study represents a junior level of AGI. With AI's reasoning capabilities already demonstrated and its potential to manage public charging operations, such as smart charging to reduce peak demand or adjusting time-of-use pricing, fully automated systems are likely to be developed in the coming years. As these systems become more prevalent, it is essential that operators or planning agencies maintain transparency about their use and decision-making processes. While human planners would still be present with such AI-automated systems, those with computational knowledge are likely to gain influence, while those without may be left behind. This will widen professional disparities and potentially lead to a new form of power imbalance in the planning process.

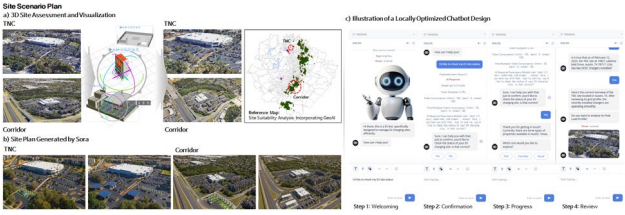

**Fig. 6.** Site Scenario Plan: (a) 3D Site Assessment and Visualization; (b) Site Plan Generated by Sora; (c) Illustration of a Locally Optimized Chatbot Design

## V. CONCLUSION: TOWARDS COMMUNITY-CENTERED URBAN PLANNING AI

This study presents multiple case studies in transportation electrification planning, including site suitability analysis and methods to enhance and manage the planning process using various AI applications and agents. It explores the use of GeoAI, GenAI, and LLMs. In summary, Urban AI offers benefits in providing and quantifying land use information. AI agents can assist planners as virtual assistants or serve as tools to reduce the time and cost traditionally required in the planning process. The development of open-source tools and software has made it easier to create 3D physical replicas of cities. However, there are limitations; empirical estimations may lack accuracy and must be evaluated through human oversight. Automated systems inherently require computationally skilled human planners at the forefront. Ultimately, the core essence of planning will remain, with human planners working in a complementary role alongside AI agents.

To mitigate biases and hallucinations in Urban AI, we should conceptually revisit how planners make decisions and how information is communicated to them. According to communicative action theory, these are distinct processes. **Figure 7** illustrates a proposed flowchart for integrating community feedback to audit Urban AI systems. Until we reach a singularity where automated Urban AI systems are both transparent and trustworthy, human planners will continue to play a leading role in informing communities and guiding planning decisions.

Of course, computational and empirical tasks, such as spatial analysis using GIS and digital twinning, will remain integral to the planning process. Planners will also need to conduct complementary activities, including User Experience (UX) testing, to evaluate the outputs of algorithmic systems. It is important to remember that planning as a discipline values recognition of care, citizen engagement, and inclusive stakeholder participation to build community knowledge and collective capacity [33]. Therefore, it is appropriate to frame this evolving area of study as Community Experience (CX) testing. This concept calls for integrating algorithmic management with deeper inquiries into algorithmic governance, while revisiting foundational theories of equity and justice [13], [34].

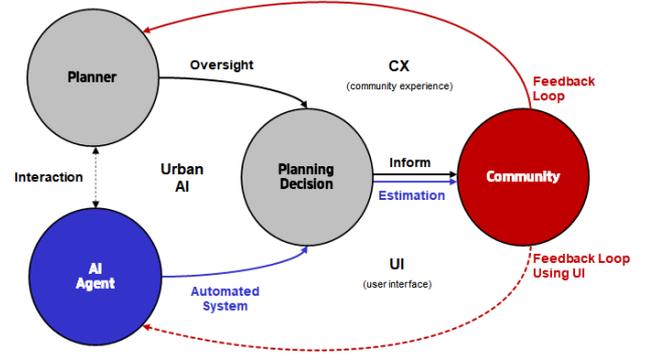

**Fig. 7.** Interaction Flow of Urban Planning AI and Community Feedback Loops

In the meantime, when automated systems make planning decisions, it should be transparently communicated to the community that the estimations are generated by AI. In cases where hallucinations or inaccuracies are suspected, communities should be provided with interfaces or tools to audit such decisions. Providing users with options to audit AI-generated decisions can enhance trust and contribute to the development of human-centered tools [35], [36]. Without such trust, users may circumvent algorithmic control or exhibit resistance [34], [37]. However, we should not expect community members to possess the same level of technical expertise as computational planners. Instead, user-friendly and human-centered User Interfaces (UIs) should be developed to support accessible auditing of AI systems. The study of such interfaces intersects with HAI, AI ethics, planning theory, and planning practice. Establishing a two-way community feedback loop in Urban AI is essential to ensure public accountability and to advance community-centered governance. I propose this direction as a critical area for future research, particularly as we move closer to the AI singularity.

Nevertheless, this study focuses solely on the application of AI in transportation electrification planning. Expectations, benefits, and potential pitfalls are likely to differ across other planning domains. The performance of the Pix2Pix GAN model could be enhanced by incorporating a larger number of samples representing distinct site types. Lastly, the chatbot demonstration does not extend to functions such as modifying power or electrical grid systems, tasks that remain beyond the current scope and warrant further investigation in future research.

ACKNOWLEDGMENT

This manuscript is a revised version of Seung Jun Choi's doctoral dissertation, completed at The University of Texas at Austin. The author declares no potential conflicts of interests with respect to the research, authorship, and/or publication of this article.
ACKNOWLEDGMENT

This manuscript is a revised version of Seung Jun Choi's doctoral dissertation, completed at The University of Texas at Austin. The author declares no potential conflicts of interests with respect to the research, authorship, and/or publication of this article.

9**APPENDIX**

**Table A**. Descriptive Statistics of Variables Used in the Baseline Site Suitability Analysis

| | variables | unit | mean | min | median | max |
|---|---|---|---|---|---|---|
| Population Density (Social Explorer, 2022) | Census population per square mile | square mile | 4,529 | 0 | 3,383 | 424,352 |
| Traffic Density (TxDoT, 2024) | Highest average annual daily traffic | score | 9,597 | 0 | 2,797 | 305,897 |
| EVCS (NREL, 2024) | Number of public EVCS: Level 2, 3 | count | 0 | 1 | 0 | 30 |
| EVCS (NREL, 2024) | Distance to the nearest non-Tesla DCFC | meter | 11,887 | 0 | 8,368 | 94,929 |
| Race (Social Explorer, 2022) | Hispanic and Black census population | % | 0.36 | 0.00 | 0.31 | 1 |
| Income (Social Explorer, 2022) | Below poverty threshold | % | 0.11 | 0.00 | 0.06 | 1 |
| Substation/Energy (OpenGridMap, 2019) | Nearest distance to the substation with at least 110 kV capacity | meter | 4,587 | 0 | 3,932 | 49,279 |
| Substation/Energy (OpenGridMap, 2019) | Nearest distance to the substation with at least 220 kV capacity | meter | 47,383 | 0 | 45,756 | 158,032 |
| Land Use (USGS, 2024) | Developed area | % | 0.65 | 0.03 | 0.70 | 1.00 |
| Land Use (USGS, 2024) | Underdeveloped area | % | 0.35 | 0.00 | 0.30 | 0.97 |
| Housing Units (Social Explorer, 2022) | Multifamily (MF) | % | 0.31 | 0.00 | 0.18 | 1.00 |
| Vehicle Ownership (Social Explorer, 2022) | Zero-vehicle households | % | 0.05 | 0.00 | 0.02 | 1.00 |
| EV Corridor (U.S. Census Bureau, 2019) | Nearest distance to EV corridor (I-35) | meter | 24,010 | 512 | 14,193 | 131,001 |
| Policy (NETL, 2022) | Disadvantaged communities as defined by Justice40 | dummy | 0.16 | 0 | 0 | 1 |